\documentclass[aip,jcp,preprint,superscriptaddress,floatfix]{revtex4-2}
\usepackage{graphicx} 
\usepackage{hyperref}
\usepackage{lineno}  
\usepackage{xcolor}
\usepackage{mhchem}
\usepackage{natmove}
\usepackage{braket}
\usepackage{cleveref}
\usepackage{amsmath}
\usepackage{amssymb}
\usepackage{tabulary}
\usepackage[shortcuts]{extdash}

\colorlet{review}{black}

\begin{document}

\title{A Density-Based Continuous Local Symmetry Measure}

\author{Duc Anh Lai}
\author{Devin A. Matthews}
\email{damatthews@smu.edu}
\affiliation{Department of Chemistry, Southern Methodist University, Dallas, TX 75275, USA}


\begin{abstract}
    Although continuous symmetry theory has attracted increasing attention in modern chemistry, local symmetry remains under-investigated. As a consequence, the relationship between symmetry and chemical behavior is often obscured, limiting the practical use of fuzzy symmetry measures. In this study, we introduce a novel framework for evaluating local symmetry based on electron density localization, and present continuous symmetry representations for several representative molecules. Our approach not only quantitatively captures global symmetry, but also reveals distinctive features of symmetry in a local chemical environment. The related concept, local chirality or chirotopicity, is also discussed. Overall, the proposed local symmetry and chirality measures provide valuable insights into molecular structure and structure-property relationships. 
\end{abstract}

\maketitle

\section{Introduction}

Symmetry and chirality are important concepts in all areas of chemistry. Yet, they are most often employed as all-or-nothing properties of a molecular system. As a consequence, we often overlook the changes of symmetry and/or chirality in a chemical transformation. A number of studies, however, indicated that the molecular symmetry varies greatly during the course of a reaction \cite{tuvi-arad_determining_2010}, due to dynamics and temperature changes \cite{tuvi-arad_effect_2013, fossepre_binding_2020}, during molecular assembly \cite{alon_continuous_2023}, and in many other scenarios. Symmetry breaking and/or formation may reveal essential insights into the reaction mechanisms, reactivity, and selectivity \cite{tuvi-arad_effect_2013, handgraaf_continuous_2005, tuvi-arad_symmetry-enthalpy_2012}. In this regard, a continuous scale for symmetry and chirality is highly desirable. Indeed, many papers related to ``continuous symmetry/chirality'' have been published recently, with foci such as structure distortions \cite{lopata_mutations_2015, neudeck_establishing_2016}, material science \cite{gakiya-teruya_546_2025}, electronic devices \cite{guo_switchable_2022}, and host-guest recognition \cite{shah_calochorturils_2025}. 

Several methods have been proposed to quantify continuous symmetry and/or chirality \cite{petitjean_chirality_2003, natarajan_numerical_nodate, zhang_methods_2012}. Such methods are not only applied in chemistry, but also span mathematics \cite{garrido_symmetry_2010}, computer sciences \cite{cham_symmetry_1995}, physics \cite{petitjean_chirality_2003}, biology \cite{marzec_morphogenesis_1999, alon_continuous_2023}, and other fields \cite{zusne_measures_1971, yodogawa_symmetropy_1982}. The oldest algorithm is Guye's asymmetry product, which is a continuous function to evaluate achirality of a set of vertices \cite{petitjean_chirality_2003, natarajan_numerical_nodate}. In 1934, a formal analog of the asymmetry product proposed by Boys was applied to estimate the optical rotatory power \cite{boys_optical_1934, boys_optical_1934-1}. Since then, generalizations of the asymmetry product have advanced theoretical models of continuous chirality, for example the chirality functions theory of Ruch and Schönhofer \cite{ruch_theorie_1970}, and the pseudoscalar quantitative chirality measure of Damhus and Schaffer \cite{damhus_three_1983}. In 1985, Gilat suggested an overlap-based symmetry measure which can be employed with masses, charges, surfaces, and wavefunctions \cite{gilat_chiral_1985, gilat_chiral_1989}. Gilat's overlap equation, which ranges from 0 to 1, is the generalized formalism that has been widely applied in modern symmetry measures. Other symmetry and chirality measures include Mezey’s quasi-symmetry from fuzzy set theory \cite{maruani_symmetry_1990, mezey_concept_1990}, molecular dissimilarity \cite{meyer_similarity_1991}, dissymmetry functions \cite{kuzmin_possible_1988}, continuous symmetry measure (CSM) and continuous chirality measure (CCM) from folding/unfolding processes \cite{zabrodsky_continuous_1992, zabrodsky_continuous_1995}, RMS chiral index \cite{petitjean_root_1999}, quantum similarity index \cite{janssens_molecular_2007}, and the helicity tensor \cite{ferrarini_assessment_1998}. 

The principal objective of such symmetry measures is to correlate with chemical properties, such as optical rotation \cite{boys_optical_1934-1}, infrared (IR) and vibrational circular dichroism (VCD) spectroscopy \cite{lipinski_local_2014}, enantiomeric excess \cite{bellarosa_enantiomeric_2003}, asymmetric catalysis \cite{alvarez_quantitative_2003}, molecular interactions \cite{keinan_quantitative_1998}, and biological activity \cite{jamroz_chirality_2012}. However, previous studies showed that such correlations are very modest, limiting the quantitative application of continuous symmetry \cite{lipinski_local_2014}. A major deficiency is the focus on calculating the total symmetry of the entire molecular system, despite the role of local symmetry of substructures. In fact, most chemical interactions and properties are associated with the local context of functional groups instead of the global level of the whole molecule. As a consequence, symmetric functional sites are often neglected due to reduced global symmetry. In addition, the need for identifying the molecular portions which are most responsible for chemical activities is more critical in rational design of molecular systems in medicinal chemistry and material engineering. On the other hand, periodic materials remain highly symmetric despite local symmetry breaking due to co-linearity cancellation, meaning that the global symmetry is artificially restricted and potentially misleading in polymer and solid-state applications. 

More importantly, the locality of stereochemistry is even more critical when it comes to processes related to molecular chirality. Mislow and Siegel discussed environment-dependent chirality, and introduced the definition of the ``chirotopicity field'' \cite{mislow_stereoisomerism_1984}. In their local chirality model, chirotopicity describes the stereochemical feature of a specific point or region in a molecule, which is regulated by the external environment, such as chiral solvents and nearby chiral interactions. This modern concept is broader than the traditional idea of chirality as it is not restricted to asymmetric carbon (stereocenter). The authors also defined chirality as an inclusive property such that all sites in a chiral molecule are sensitive to chirality (chirotopic), while sites in an achiral molecule can be either achirotopic or chirotopic. Furthermore, several studies of VCD signatures found a significant effect of local characteristics, such as intramolecular hydrogen bonds \cite{mazzeo_spectroscopic_2022}. 

To address the importance of local symmetry and chirality theory, we propose a novel approach to quantify local symmetry of a molecular system based on the local projection and transformation of the electronic density. Subsequently, we present symmetry measures of several molecules in terms of different symmetry operations. Although local symmetry is our focus, molecular symmetry from the local to the global level can be smoothly probed by control over the radial extent of the local projection basis. Overall, the global symmetry calculations are precisely consistent with the traditional assignment, confirming the generalized performance of our method. Whereas, the local symmetry measures are descriptive and complementary. 

\section{Theory}\label{sec:theory}
In general, quantitative symmetry methods evaluate the relationship between a molecular object and its symmetry-related counterpart, and can be represented by either a distance normalization metric, for example, 
\begin{equation}
    S(\tau)\propto\Vert O-\hat O(\tau)\Vert
    \label{eq:S}
\end{equation}
or an overlap metric,
\begin{equation}
    S(\tau)\propto\langle O|\hat O(\tau)\rangle
    \label{eq:S2}
\end{equation}
where $O$ is a representation of the original molecule, such as atomic coordinates \cite{zabrodsky_continuous_1992}, molecular volume \cite{gilat_chiral_1989}, topology \cite{zhang_methods_2012}, or electron configuration \cite{grimme_continuous_1998, bellarosa_enantiomeric_2003}, and $\hat O$ is the corresponding symmetry-related quantity, often with respect to a particular symmetry operation $\tau$. Based on the essence of $\hat O$, existing continuous symmetry measures can be classified into two main categories \cite{petitjean_chirality_2003}. The first category assesses the similarity between an object and its image generated by a symmetry operation. These approaches require the calculations to be minimized for all for all rotations and translations to find the optimal symmetry operator on a given molecule. Consequently, iterative optimizations are typically involved in calculations of the first approach. The second category quantifies the deviation of the object from an idealized symmetric reference. This approach is generally more computational efficient because ones only need to carry out a single similarity measure once the reference is chosen. Nevertheless, the choice of reference structure is inherently system dependent, which biases and complicates comparisons between symmetry measures of different molecules. 

When $O$ and $\hat O$ are localized representations of a molecule, the metric $S$ can be used to quantify local symmetry and/or chirality. Alvarez et al. adapted this fundamental idea to measure local symmetry of coordination complexes. The authors partitioned transition metal complexes into successive shells and introduced the concept of shell chirality to extract chirality relationships between the coordination sphere, the ligands, and the whole molecule \cite{alvarez_quantitative_2003}. Subsequently, Lipinski and co-workers applied a similar idea to quantify local chirality from small fragments of molecules, demonstrating that local measures on chirality of truncated structures correlate with IR and VCD spectroscopic features \cite{lipinski_local_2014}. Moreau followed the same approach to come up with an assessment of atomic chirality based on principal planes defined by the environment of an atom \cite{moreau_atomic_1997}. However, these approaches do not fully integrate electronic coupling with neighboring atoms (or molecules in the case of intermolecular interactions), and therefore do not satisfy additivity and transferability. An alternative strategy is to derive $O$ and $\hat O$ from the electron density \cite{janssens_molecular_2007}. The density offers two-fold advantages: first, the density continuously spans the entire space, which ensures $S$ will always be a continuous function, and second, the density can be smoothly partitioned into contributions of an atom, group of atoms, or spatial region from which local properties can be derived. Moreover, the electronic density incorporates environmental and electronic effects that are absent from purely geometric representations. For instance, perturbations induced by solvents or external fields often manifest more strongly in the electron density than in nuclear coordinates. In addition, electronic structure plays a central role in resonance and charge-transfer phenomena, making density-based symmetry measures particularly suited for chemically relevant symmetry evaluation. 

We can obtain the local density matrix at a point $A$ by projecting the total density matrix to a local basis set centered at $A$, and use this local density to compute $S$. From the spirit of \Cref{eq:S}, the local continuous symmetry at point $A$ can be formulated as
\begin{equation}
    S_A(\tau)=1-\frac{\Vert\mathbf{D}_A-\mathbf{\hat D}_A(\tau)\Vert_F}{\Vert\mathbf{D}_A\Vert_F+\Vert\mathbf{\hat D}_A(\tau)\Vert_F}
    \label{eq:Sa}
\end{equation}
where $\mathbf D_A$ denotes the local density matrix at point $A$, and $\mathbf{\hat D}_A$ is the density matrix after applying the symmetry operation $\tau$ on the local density matrix. The density localization is obtained by a projection from the original (global) basis set $\mathcal{B}$ to a local, point-centered basis set $\mathcal{B}_A$,
\begin{align}
    \mathbf{D}_A&=\mathbf S_{AA}^{-1}\mathbf S_{GA}^T\mathbf{D}\mathbf S_{GA}\mathbf S_{AA}^{-1}
    \label{eq:density projection}
\end{align}
where $\mathbf D$ is the one-electron reduced density matrix (1RDM) in the global basis set, and $\mathbf S_{GA}$ and $\mathbf S_{AA}$ are the overlap matrix between global and local basis sets and the overlap matrix of the local basis set, respectively. For common Gaussian-type orbital (GTO) basis sets, we can achieve the inverse overlap, $\mathbf S_{AA}^{-1}$, by using orthogonalization transformations, or Cholesky decomposition of the non-diagonal overlap $\mathbf S_{AA}$. 

One of the advantages of projecting the density matrix onto a point-centered basis set is that it is closed under operations defined by points, axes, or planes passing through that point, making application of symmetry elements trivial. We employ an uncontracted GTO basis set in the Cartesian angular momentum representation,
\begin{align}
    \mathcal{B}_A&=\{\phi^{abc}_A(r)\,|\,0 \le l = a+b+c \le l_{max}\}
    \label{eq:basis function} \\ 
    \phi^{abc}_A(r)&=\mathcal{N}(x-A_x)^a(y-A_y)^b(z-A_z)^ce^{-\zeta(l,R)|r-A|^2} \\
    \zeta(l,R)&=\frac{1}{2\pi}\left(\frac{(2l+2)!!}{(2l+1)!!R}\right)^2
\end{align}
where $\mathcal{N}$ is a normalization factor and $l_{max}$ and $R$ are the user-specified maximum angular momentum and average radial extent, respectively. The orientation of the function is expressed through the powers of $x,y,z$, with the exponents $a,b,c$ defining the Cartesian angular momentum of the basis function. Results from an application of a symmetry operator $\tau$ on a basis function can be expressed as,
\begin{equation}
    \tau\phi^{abc}_A=\sum_{a'+b'+c'=a+b+c}P_l(\tau)^{abc}_{a'b'c'}\phi^{a'b'c'}_A
\end{equation}
\textcolor{review}{where the summation runs over all possible combinations of $\{a',b',c'\}$ for each angular subshell $l=a+b+c$, and the transformation coefficients $P_l(\tau)$ can be easily constructed from the $3\times3$ matrix $P_1(\tau)$ defining the symmetry operation in $\mathbb{R}^3$.}
A square transformation matrix $\mathbf {P}(\tau)$ for the entire local basis is then constructed from the transformation coefficients for each $l$. 
The symmetry-related density matrix $\mathbf{\hat D}_A(\tau)$ is therefore obtained as,
\begin{equation}
    \mathbf{\hat D}_A(\tau)=\mathbf{P}(\tau)\mathbf{D}_A\mathbf P(\tau)^T
\end{equation}
In \Cref{eq:Sa}, the quantitative symmetry is normalized by a sum of Frobenius norms of $\mathbf D_A$ and $\mathbf {\hat D}_A(\tau)$ in the denominator, resulting in $S_A$ ranging from 0 to 1. $S_A$ reaches the upper limit only if $\mathbf D_A=\mathbf{\hat D}_A(\tau)$. In other words, the local electron density is invariant with respect to the symmetry operation, and the molecule possesses the symmetry at the considered point $A$. Lower values of $S_A$ represent more distortion in the electron density after the symmetry operation, and hence exhibit less symmetry. For instance, if $S_A>S_B$, we can say that the electron density at point $A$ is more symmetric than that at point $B$. To extend this analysis to chirality, we also define the chirality measure as,
\begin{align}
C_A=1-\max_{n=1}^{\infty} S_A(S_{n})\label{eq:chiral}
\end{align}
In practice, only a finite number of improper rotation axes can be checked; in this work we consider operations up to $S_8$.

Although any basis set, in principle, can be directly used as a local basis for density projection (\Cref{eq:density projection}), in this work we focus on ab-initio molecular calculations, which commonly employ GTO functions. 
\textcolor{review}{Conventional GTO basis sets are systematically designed to reproduce the behavior of Slater type orbitals, particularly in the near-nucleus and asymptotic regions.}
The spatial completeness is obtained by contracting additional radial functions, $e^{-\zeta r^2}$, with different exponents, such as diffuse components. This strategy, however, introduces wide ranges of spatial distribution, thereby systematically breaking down the locality of the basis as its size increases. On the other hand, if one restricts the basis to remain compact and simple, the truncation of angular momentum shells limits its ability to represent anisotropic features of the electron density, such as polarization and directional bonding, which provide symmetry insights, as well as angular resolution of more distant portions of the electronic density. Even in large basis sets, the maximum angular momentum quantum number remains relatively low, while the addition of diffuse Gaussian functions leads to a significant increase in radial extent. To address this imbalance and retain the spatial locality, we suggest a new type of local basis set. This basis set employs an uncontracted GTO formalism (\Cref{eq:basis function}), but with Gaussian exponents adjusted such that all basis functions hold the same expectation value of the radial coordinate $\langle r\rangle\equiv R$, enforcing a uniform spatial locality across different angular momentum shells (Figure~S1). In addition, we omit the contraction of multiple functions, instead each angular momentum shell is represented by a single primitive GTO, and include angular momenta up to $g$ functions ($l=4$), yielding a total of 35 Cartesian GTOs. As a result, the tailored basis set provides a balance between radial locality and angular completeness for the local density representation.
\textcolor{review}{The proposed symmetry measure \Cref{eq:S} is therefore considered as a spatially localized functional of the one-particle density matrix projected onto a finite region defined by the parameter $R$. In this regard, the measure can be viewed as a coarse-grained descriptor of how the electronic density within a given spatial domain transforms under a chosen symmetry operation. Therefore, the present approach enables identification of spatial regions where symmetry is locally preserved or broken, even when the overall molecular structure belongs to a high-symmetry group. From a physical perspective, local symmetry breaking in the electron density is directly relevant to chemical behavior. Regions of reduced symmetry often coincide with anisotropic charge distributions, directional bonding environments, and sites of enhanced chemical reactivity or intermolecular interaction. In this way, the proposed measure provides a bridge between abstract symmetry operations and chemically meaningful concepts such as functional groups, binding sites, and stereochemically active regions.}

\begin{figure}
    \centering
    \includegraphics[width=\linewidth]{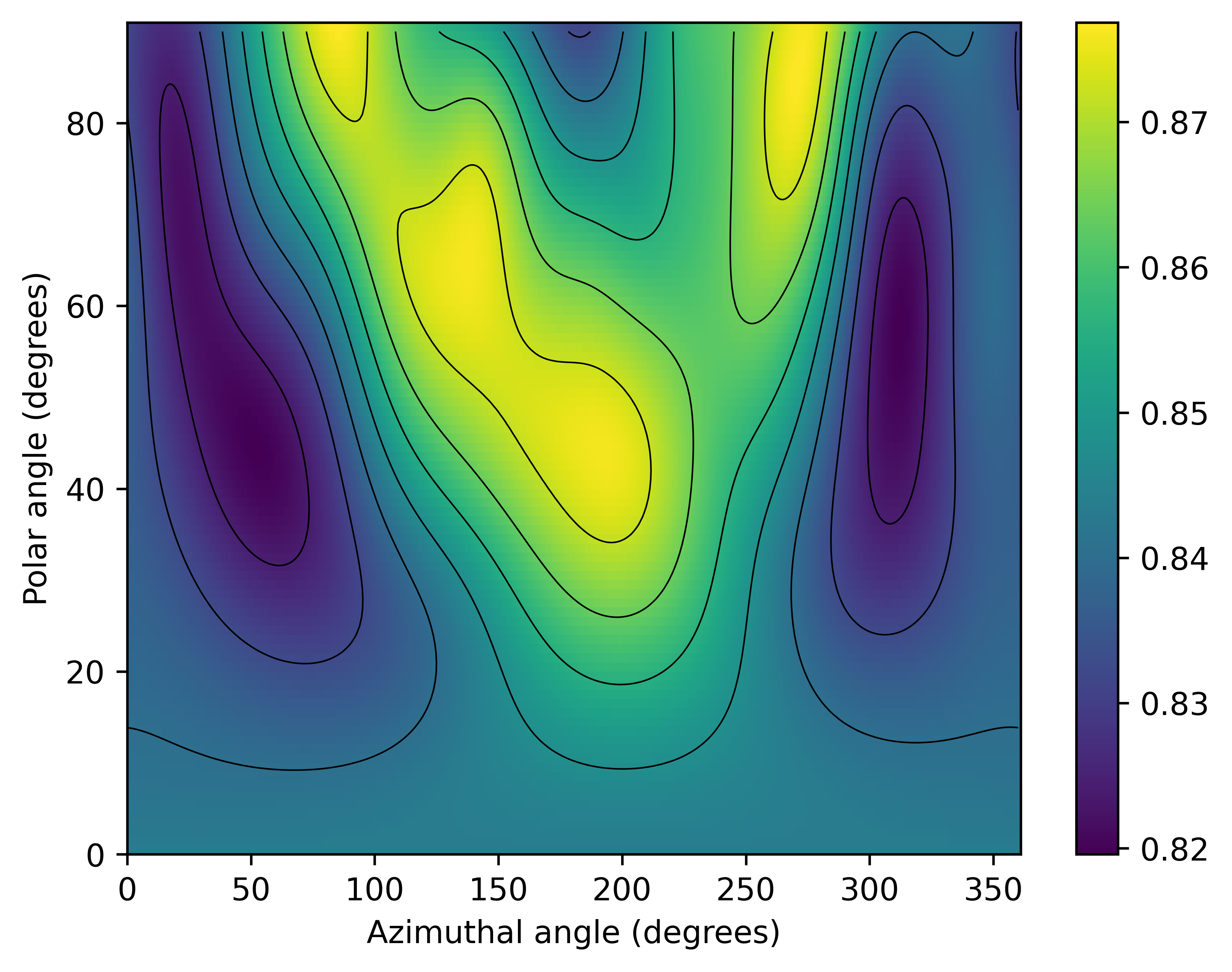}
    \caption{Contour plot of local plane symmetry at the $\alpha$ carbon of 1-pentanol [$S_{\ce{C_{\alpha}}}(\sigma)$] as a function of polar and azimuthal angles of the $\sigma$ plane.}
    \label{fig:surface}
\end{figure}

Given a predefined symmetry operation, we can compute the symmetry-transformed local density $\mathbf{\hat D}$, and the local symmetry measure is evaluated using \Cref{eq:Sa}. If the objective is to determine the optimal symmetry measure associated with a given symmetry element, an optimization over all possible symmetry operators is required. The goal of the optimization is to determine parameters of the symmetry operator, such as the direction of the rotation axis, the normal of a reflection plane, or both in the case of improper rotation, that maximize the local symmetry measure. In the present method, the optimization is performed with respect to the azimuthal and polar angles in a spherical coordinate representation. As shown in \Cref{fig:surface}, the reflection symmetry measure of 1-pentanol computed at the $\alpha$ carbon exhibits a highly complex landscape with multiple flat maxima, indicating that the choice of optimization algorithm is critical. In the current implementation, we employ a brute-force grid-based search to explore the parameter space, increasing the chance of locating the global optimum, but at the cost of significant computational overhead. To provide additional flexibility, the current implementation also allows users to supply external optimization routines. However, such optimizers should be used with caution, for example, stochastic optimization strategies may fail to converge to the true global optimum in the presence of multiple shallow maxima. 

\section{Implementation}

The Python implementation of the continuous local symmetry measure is available on Github\cite{git}. The overlap matrices are computed using the PySCF package \cite{sun_pyscf_2018, sun_recent_2020}. Normally, a calculation starts with a PySCF molecule instance, and its corresponding density matrix. However, total density matrix can also be read from the MOLDEN \cite{schaftenaarMolden20Quantum2017} output of other quantum chemistry suites, e.g. Q-Chem \cite{epifanovskySoftwareFrontiersQuantum2021}, ORCA \cite{neese_software_2025}, CFOUR \cite{matthewsCoupledclusterTechniquesComputational2020}, and others with a helper function called \texttt{read\_molden}. All symmetry operations, including identity, rotation, reflection, inversion, and improper rotation, are supported under the \texttt{sym\_op} submodule. Radial extent is controlled through the \texttt{radius} argument, which is specified in atomic units. As mentioned at the end of \Cref{sec:theory}, users can provide an external optimizer function through the \texttt{optimizer} argument, as long as it accepts the same arguments and returns equivalent results as built-in SciPy \cite{virtanenSciPy10Fundamental2020} optimizers. In this work, all densities are calculated at the B3LYP/6-31G(d,p) level, and with the exception of 1-pentanol (which uses a synthetic geometry), all geometries are optimized at the same level.

\section{Results and Discussion}

\subsection{Local Symmetry}

First we examine how a single chemical substituent, e.g. a hydroxyl group (-OH), perturbs local symmetry, using 1-pentanol as a prototypical example. Without the hydroxyl group, $n$-pentane contains a $\sigma$ reflection plane coincident with the plane of the carbon backbone. As a consequence, any point lying on this plane must measure perfect local mirror symmetry (i.e. $S_A(\sigma)=1\,\forall\,A\in\sigma$), in particular at the carbon centers themselves. However, introducing an out-of-plane hydroxyl substituent at the terminal carbon breaks this mirror symmetry in the global context. It should be noted that our analysis considers a static, idealized conformation of 1-pentanol that lacks mirror symmetry but maintains the planar carbon backbone. In practice, the observable structure of 1-pentanol adopts effective $C_s$ symmetry due to the thermally-accessible \ce{C_{\alpha}-C_{\beta}} rotation.

\begin{figure}
    \centering
    \includegraphics[width=\linewidth]{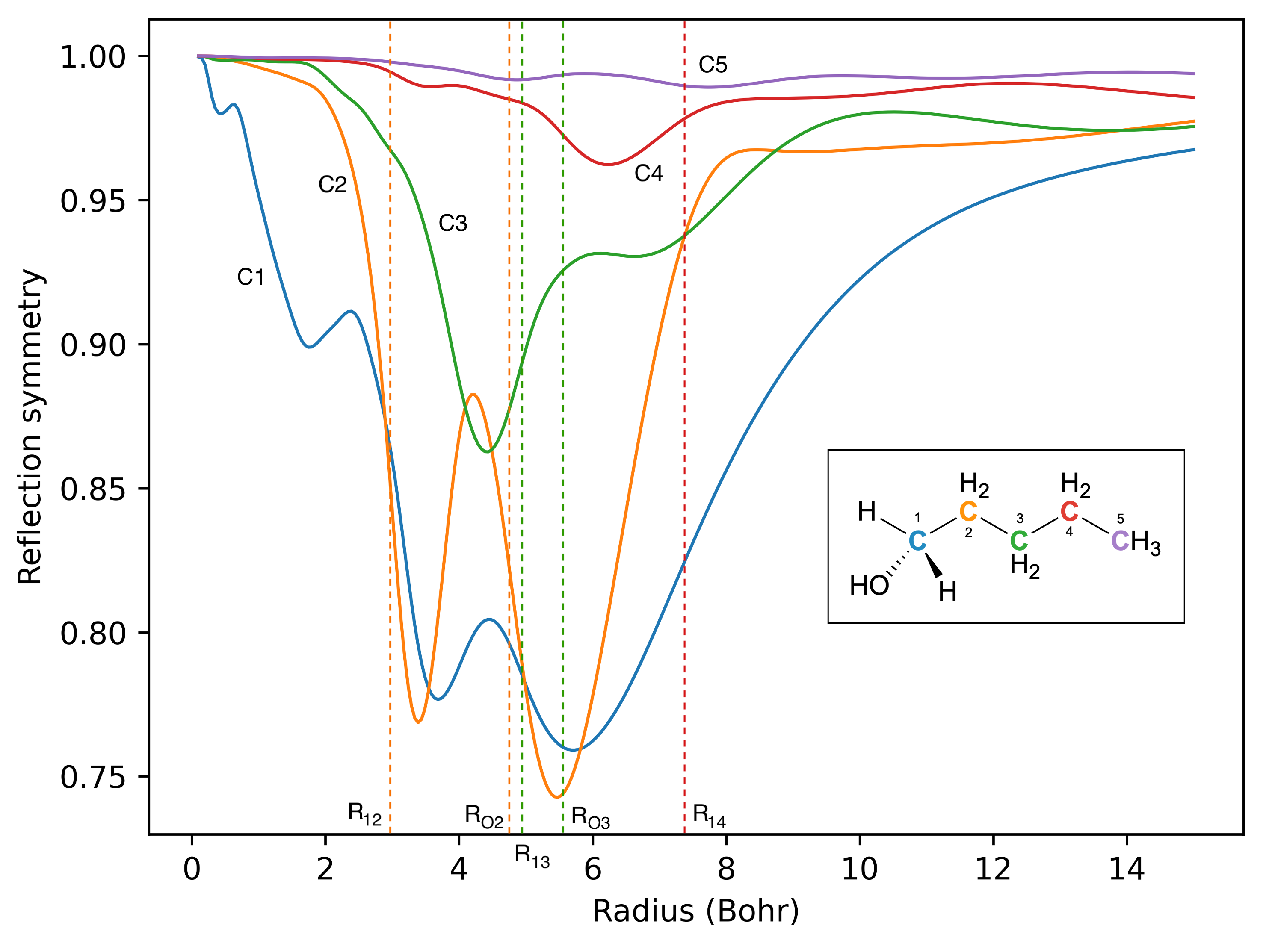}
    \caption{Local reflection symmetry at carbon centers, $S_{\text{C}n}(\sigma)$, computed with varying radial extent $R$. Vertical dashed lines indicate selected internuclear distances.}
    \label{fig:pentanol}
\end{figure}

\begin{figure}
    \centering
    \includegraphics[width=\linewidth]{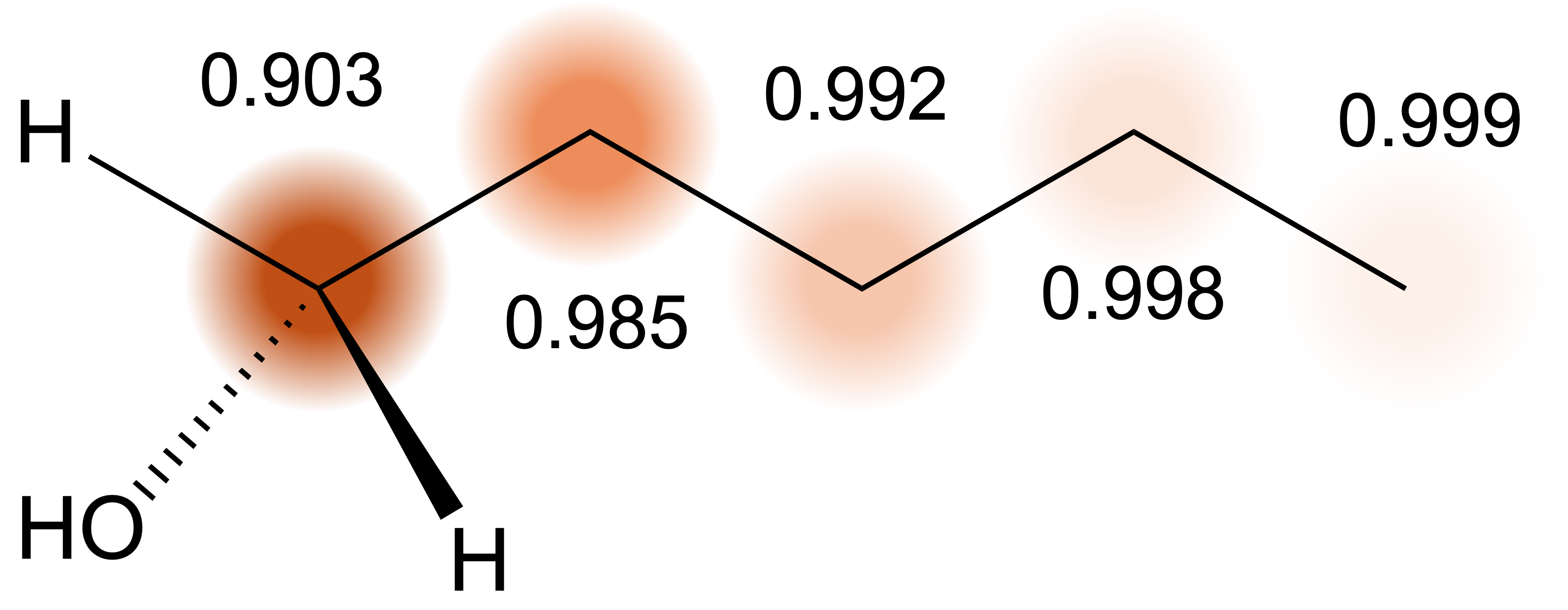}
    \caption{Local reflection symmetry at carbon centers, $S_{\text{C}n}(\sigma)$, with $R$ fixed at 2 a.u.}
    \label{fig:pentanol2}
\end{figure}

In this analysis, we demonstrate that different segments of the molecule experience the symmetry distortion induced by the asymmetric functional group (e.g. \ce{-OH}), with differing but explainable magnitudes based on distance from the substitution center and radial extent of the local basis set. The local reflection symmetries computed at each carbon atom with various radial extents are presented in \Cref{fig:pentanol}. An asymptotic behavior is observed for all carbon centers as the symmetry measures approach 1.0 for extremely small and large radii. Whereas, meaningful values of local symmetry are obtained at finite values of $R$, although the meaning varies with radial extent, as observed below. In the example of 1-pentanol, the local symmetries drop significantly as the radius increases. The lowest mirror symmetry is 0.743 and occurs at C2, indicating the strongest (semi-)local perturbation, with an average radial extent of 5.43 a.u. The magnitudes of symmetry distortions are largest on C1 and C2, which are close to the hydroxyl group, and decay significantly in further region such as C4 and C5. For example, the symmetry deformation at C5 is negligible compared to that at other proximal carbons. In the local atomic region, e.g., $R < 3$ a.u., the local reflection symmetries follow the order: $S_{C5}>S_{C4}>S_{C3}>S_{C2}>S_{C1}$, obviously demonstrating that the influence of the substituent on local symmetry diminishes with distance. For instance, the local reflection symmetry defects, measured as $1-S_{\text{C}n}(\sigma)$ at $R=2$ a.u. decrease exponentially from C1 to C5 (\Cref{fig:pentanol2}). The spatial extent of the local probe basis (quantified by the average radial extent $R$) has a significant effect on the computed symmetry measures, and provides insight into the spatial relationships between local density regions and density sub-structure. First, C1 is the only center which exhibits significant symmetry deviation in the local atomic region. Here, the symmetry measure clearly reveals how the core and valence shells are separately affected by the symmetry perturbation. Then, the interatomic distances highlight radial extents with high symmetry deviation. For example, the symmetry at C3 and C4 results in peaks which slightly precede the corresponding C1--C$n$ distances. Instead, the peaks for C1 and C2 lag slightly behind the \ce{C-C} distances, or almost equivalently the C1--O distance in first interatomic peak of C1. The second peak in the symmetry deviation of C3, which occurs nears 6.7 a.u. (3.5 \AA) does not clearly align with any \ce{C-C} or \ce{C-O} distance, but may instead probe the asymmetric density of the alcohol hydrogen. As there is essentially no corresponding density at the reflection image position, this should cause a significant loss of symmetry.

\textcolor{review}{It should be also noted that the parameter $R$ defines the spatial extent of the density projection (\Cref{eq:density projection}) and thus controls the length scale over which symmetry is evaluated. Smaller values of $R$ probe highly localized features of the electron density, while larger values incorporate progressively more delocalized contributions. The symmetry breaking at very large $R$ is attenuated, reflecting the finite size of the molecular system compared to the sampling ``volume'', rather than recovering the true global symmetry of the molecule.}


\begin{figure}
    \centering
    \includegraphics[width=\linewidth]{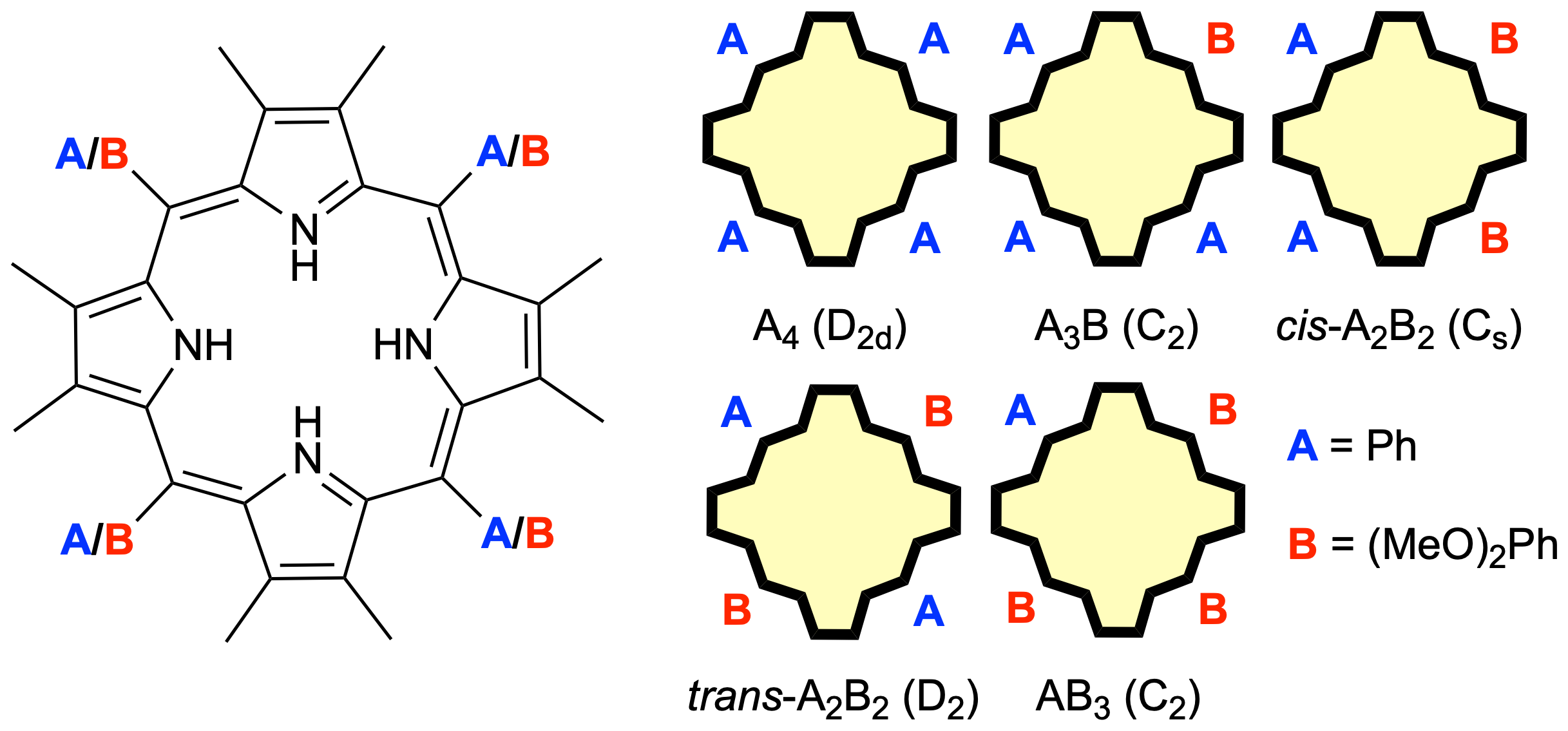}
    \caption{Diprotonated octamethyltetraphenylporphyrin (\ce{H2[OMTPP]}) framework and dimethoxyphenyl\-/substituted analogues.}
    \label{fig:porphyrins}
\end{figure}

\begin{figure}
    \centering
    \includegraphics[width=\linewidth]{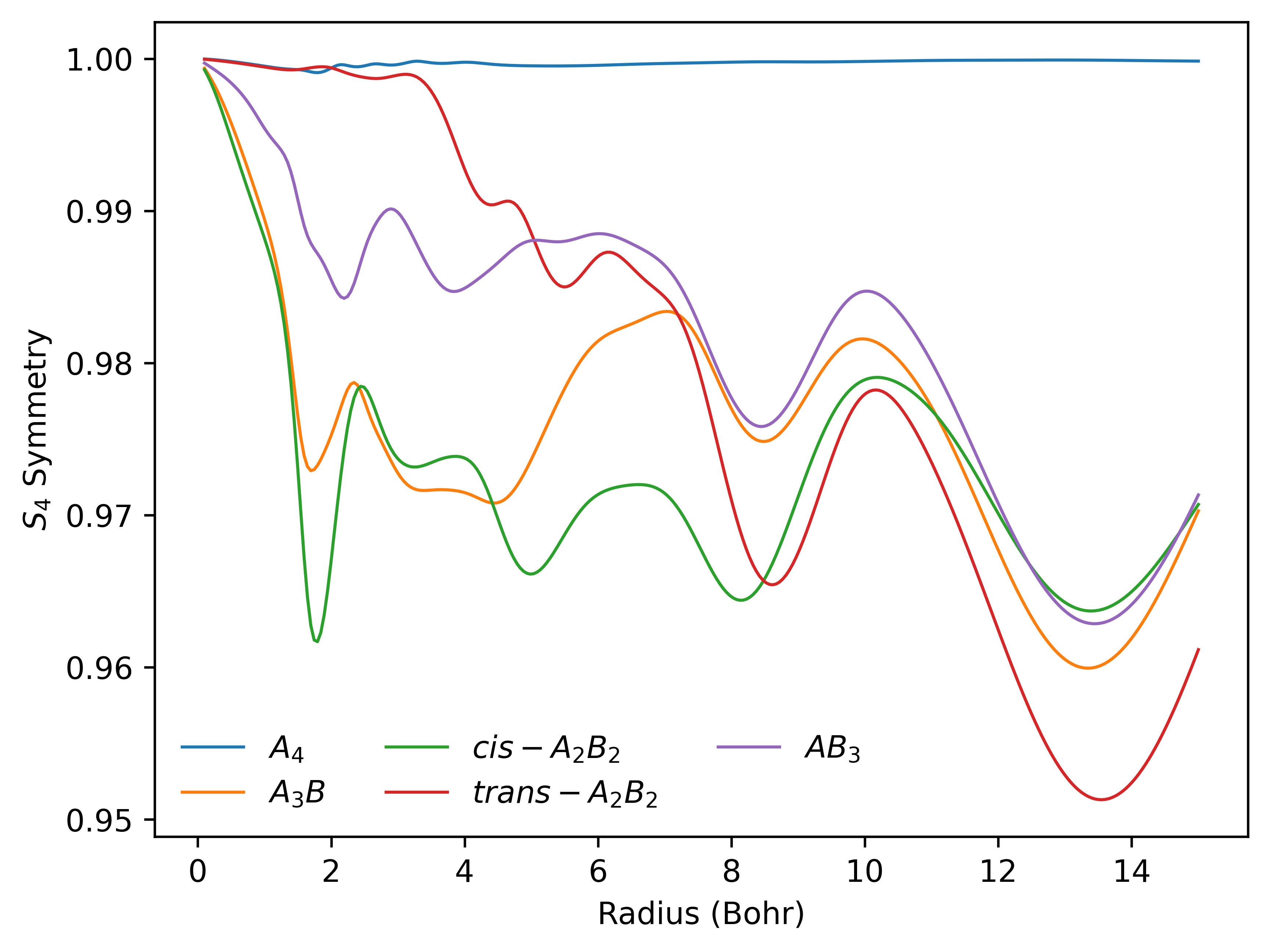}
    \caption{Local $S_4$ symmetry at the molecular center, $S_O(S_4)$, computed with varying radial extent $R$.}
    \label{fig:s4-porphyrins}
\end{figure}

As a second example, we illustrate how the probing radius can be used to quantify the symmetry from local to global scale. Porphyrins are incredibly abundant and crucial in photocatalysis, material framework, medicine, and electronic devices. In this manuscript, we study the $D_{2d}$ symmetry of diprotonated octamethyltetraphenylporphyrin (H$_2$[OMTPP]) which belongs to the A$_4$ category (\Cref{fig:porphyrins}). Specifically, H$_2$[OMTPP] possesses an $S_4$ improper rotation axis and a $C_2$ rotation axis perpendicular to the porphyrin plane. The remaining two rotation axes lie along opposite \textit{meso}-phenyl groups. Each reflection plane bisects a pair of facing pyrrole units. Gradually replacing \textit{meso}-phenyls with dimethoxyphenyls reduces the symmetry of A$_4$ porphyrin. In particular, mono-substituted A$_3$B only presents a $C_2$ rotation axis passing through the porphyrin center and the dimethoxyphenyl. Likewise, di-substitution at consecutive \textit{meso} positions, \textit{cis}-A$_2$B$_2$, results in a $C_s$ molecule, in which only the mirror plane separating the two substituents remains. In contrast, di-substituting with dimethoxyphenyls at two opposing \textit{meso}-positions, \textit{trans}-A$_2$B$_2$, preserves all three rotation axes, yielding a $D_2$ symmetric molecule. Finally AB$_3$, bearing three dimethoxyphenyl modifications on the porphyrin ring, retains only a single $C_2$ rotation axis aligned with the unsubstituted phenyl. Due to the symmetry breaking, A$_3$B, \textit{cis}-A$_2$B$_2$, \textit{trans}-A$_2$B$_2$, and AB$_3$ porphyrins exhibit atypical optoelectronic properties \cite{holzel_addressing_2021} and chiral preferences \cite{mizuno_chirality-memory_2000}.

Numerous studies have showed that the central cavity of the macrocycle serves as the functional site of porphyrins \cite{hiroto_synthesis_2017}. As a consequence, any perturbation to the central cavity can considerably modify the chemical properties and reactivity of porphyrins. The symmetry measures with respect to the electron density at the centroid of tetrapyrroles reveal essential characteristics of these perturbations induced by the modifications on the periphery of the porphyrin series, e.g., four-fold improper rotation symmetry (\Cref{fig:s4-porphyrins}). Highly symmetrical A$_4$ porphyrin exhibits a constant, ideal $S_4$ symmetry value across all radii (apart from small numerical artifacts from the grid search procedure). On the contrary, the presence of substituents generates a more complex electron density at the ring centroid, which is effectively indicated in the symmetry measures.
\textcolor{review}{Importantly, the spatial distribution of the symmetry measure can be directly related to experimentally relevant phenomena. For example, in chiral systems, regions exhibiting strong local chirotopicity are expected to correspond to domains that contribute most significantly to enantioselective interactions, such as differential binding. 
}

\begin{table}[]
    \centering
    \begin{tabular}{|c|c|c|c|c|c|}
    \hline
       Operation  &\quad A$_4$ \quad\, & \quad A$_3$B \quad\, & \textit{cis}-A$_2$B$_2$ & \textit{trans}-A$_2$B$_2$ & \quad AB$_3$ \quad\, \\
       \hline 
        $S_4$(z) & 1.000 &	0.961	& 0.964	& 0.953 &	0.964 \\
        $C_2$(z) & 1.000	& 0.955	& 0.950	& 1.000	& 0.959 \\
        $C_2'$ & 1.000	& 1.000	& 0.964	& 1.000	& 0.959 \\
        $C_2''$ & 1.000	& 0.955	& 0.964	& 1.000 &	1.000 \\
        $\sigma_d$ & 1.000	& 0.961	& 1.000 &	0.953 &	0.964 \\
        $\sigma_d'$ & 1.000	& 0.961	& 0.950	& 0.953	 & 0.964 \\
        \hline
    \end{tabular}
    \caption{Global symmetry values for all operations in $D_{2d}$, evaluated at the porphyrin center with $R=13$ a.u.}
    \label{tab:sym-porphyrins}
\end{table}

In the core porphyrin region, e.g. $R < 9$ a.u., the symmetry measures for all but two porphyrins fluctuate strongly, revealing local symmetry breaking at the porphyrin center. A$_3$B and \textit{cis}-A$_2$B$_2$ porphyrins display the largest deviations, whereas \textit{trans}-A$_2$B$_2$ present a minimal perturbation. Previous studies reported that A$_3$B, \textit{trans}-A$_2$B$_2$, and AB$_3$ porphyrins act as molecular sensors for chiral carboxylic acids \cite{mizuno_chirality-memory_2000}.
\textcolor{review}{
Chiral recognition between porphyrins and (\emph{R})/(\emph{S})-mandelic acid originates from a combination of a strong hydrogen bond between the carboxylate group and the protonated pyrrole units at the central site, together with weaker, nonspecific non-covalent interactions (NCIs) in the peripheral regions \cite{mizuno_chirality-memory_2000}. Differences in these interactions between the two enantiomers give rise to the observed chiral preference (stronger binding/stabilization of the ``matching'' enantiomer compared with the other). Within this framework, the pronounced variations in the $S_4$ symmetry measure at the central site indicate a significant local symmetry breaking in the region responsible for hydrogen bonding and partially explain their chiral recognition capabilities.}
On the other hand, \textit{trans}-A$_2$B$_2$ exhibits only slight symmetry breaking in the pyrrole ring region and essentially no symmetry breaking in the active site ($R \lesssim 4$ a.u.). 
\textcolor{review}{This observation indicates that a more in-depth analysis is required to explain chiral recognition in \textit{trans}-A$_2$B$_2$, and potentially that changes in hydrogen bonding are not the main source of chiral recognition, particularly as \textit{trans}-A$_2$B$_2$ is experimentally determined as having the strongest chiral recognition among these systems \cite{lai2026continuouslocalsymmetryconnection,mizuno_chirality-memory_2000}.}
Despite substantial differences in the local regime, the symmetry measures converge to approximately the same values when the probing radius exceeds 13 a.u., encompassing the entire molecule. Notably, the symmetry measure of A$_3$B resembles that of \textit{cis}-A$_2$B$_2$ at small radii, but gradually approaches the behavior of AB$_3$ at larger radii. 

\Cref{tab:sym-porphyrins} summarizes the global measures computed with $R = 13$ a.u. for all symmetry operators in the $D_{2d}$ point group. Substitution at the \emph{meso} positions on the porphyrin ring slightly alters the total symmetry. For all molecules, the algorithm correctly identifies the extant symmetry elements, and equivalent symmetry operations yield identical measures. For example, in A$_3$B porphyrin, the $S_4$ produces the equivalent configuration as applying the reflections, as indicated by the same symmetry measures. Similar equivalences are observed in other porphyrins, validating the consistency of the symmetry analysis.

\begin{figure}
    \centering
    \includegraphics[width=\linewidth]{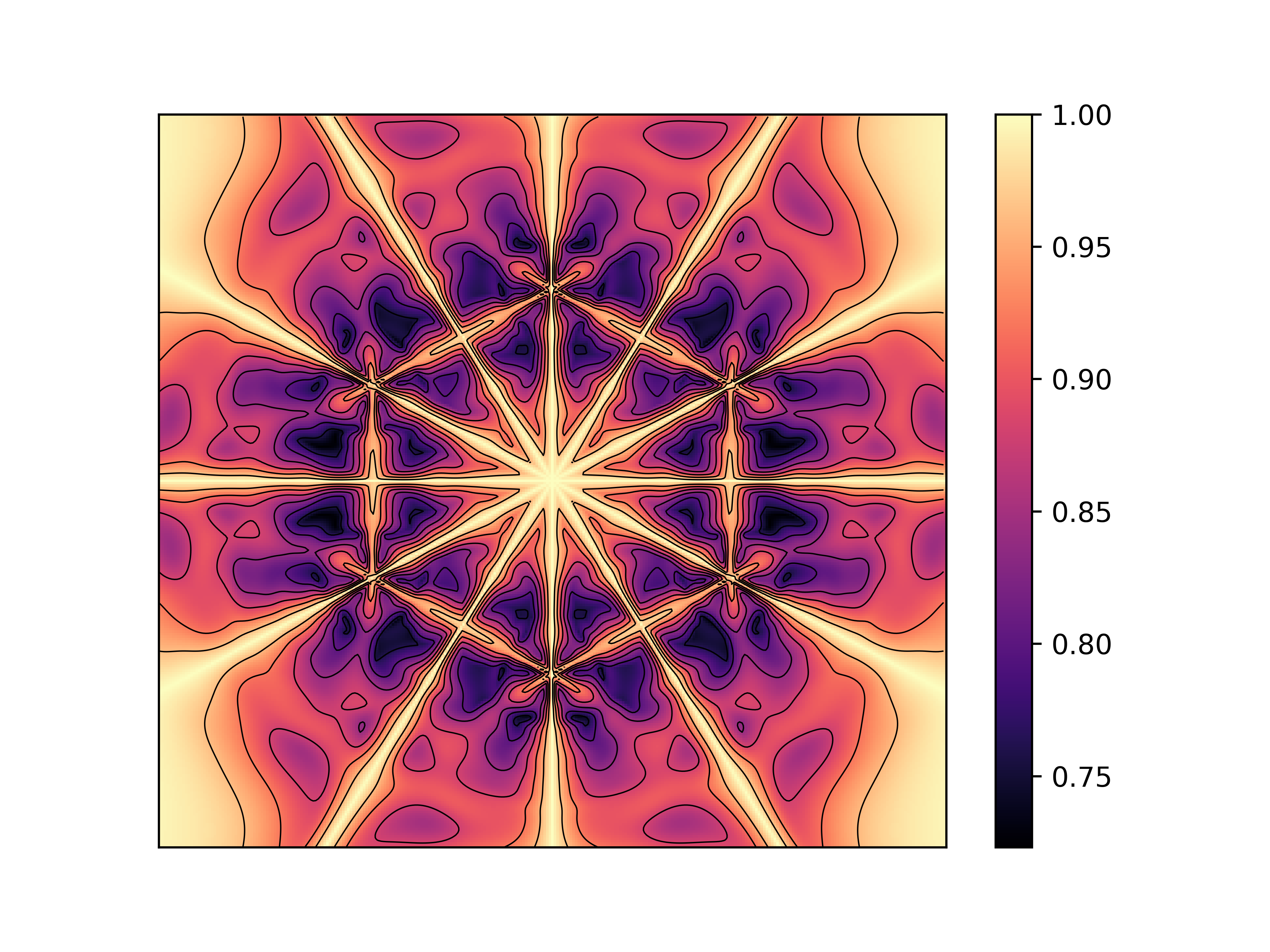}
    \caption{$C_2$ symmetry field of benzene, measured in the molecular plane with $R=1$ a.u. The position of the carbons can be inferred from the intersecting (quasi-)symmetry axes, while the hydrogens are located around the faint ring of symmetry enhancement.}
    \label{fig:benzene}
\end{figure}

As a final example, we examine the continuous local symmetry field of benzene and its fluoro-substituted species. Unsubstituted benzene is stabilized in the planar conformation and characterized by a high degree of symmetry including a principal $C_6$ rotation axis (through the ring center, perpendicular to the molecular plane), six $C_2$ axes perpendicular to $C_6$, a horizontal mirror plane (containing the ring), six vertical mirror planes normal along $C_2$ axes, and an inversion point at the center (i.e. $D_{6h}$). The abundance of multiple rotational and reflectional symmetries makes it one of the most symmetric chemical structures commonly encountered in organic chemistry. In this analysis, we focus on local $C_2$ rotations and how these symmetries collapse due to fluorine substitutions. 
\textcolor{review}{All symmetry calculations were carried out on a dense uniform grid with $\sim0.015$ \AA\ spacing on the benzene plane. The close spacing of the sampling grid reduces qualitative dependence of the resulting symmetry field w.r.t. radial extent $R$; we choose a moderate value $R=1$ a.u. which balances locality and numerical stability.}
As shown in \Cref{fig:benzene}, the $C_2$ symmetry features of benzene are obviously represented as lines of perfect symmetry aligning with the three $C_2$ axes passing through carbon atoms, and three $C_2$ axes bisecting the carbon-carbon bonds.
Approximate $C_2$ symmetry also presents along \ce{C-C} bonds due to the roughly cylindrical distribution of $\sigma$ electrons and above/below plane symmetry of $\pi$ electrons, which is only correct in terms of  a local model (note that the negative character of the $\pi$ electrons under $C_2$ is removed in the density, as is reflects the square of the wavefunction).

\begin{figure}
    \centering
    \includegraphics[width=\linewidth]{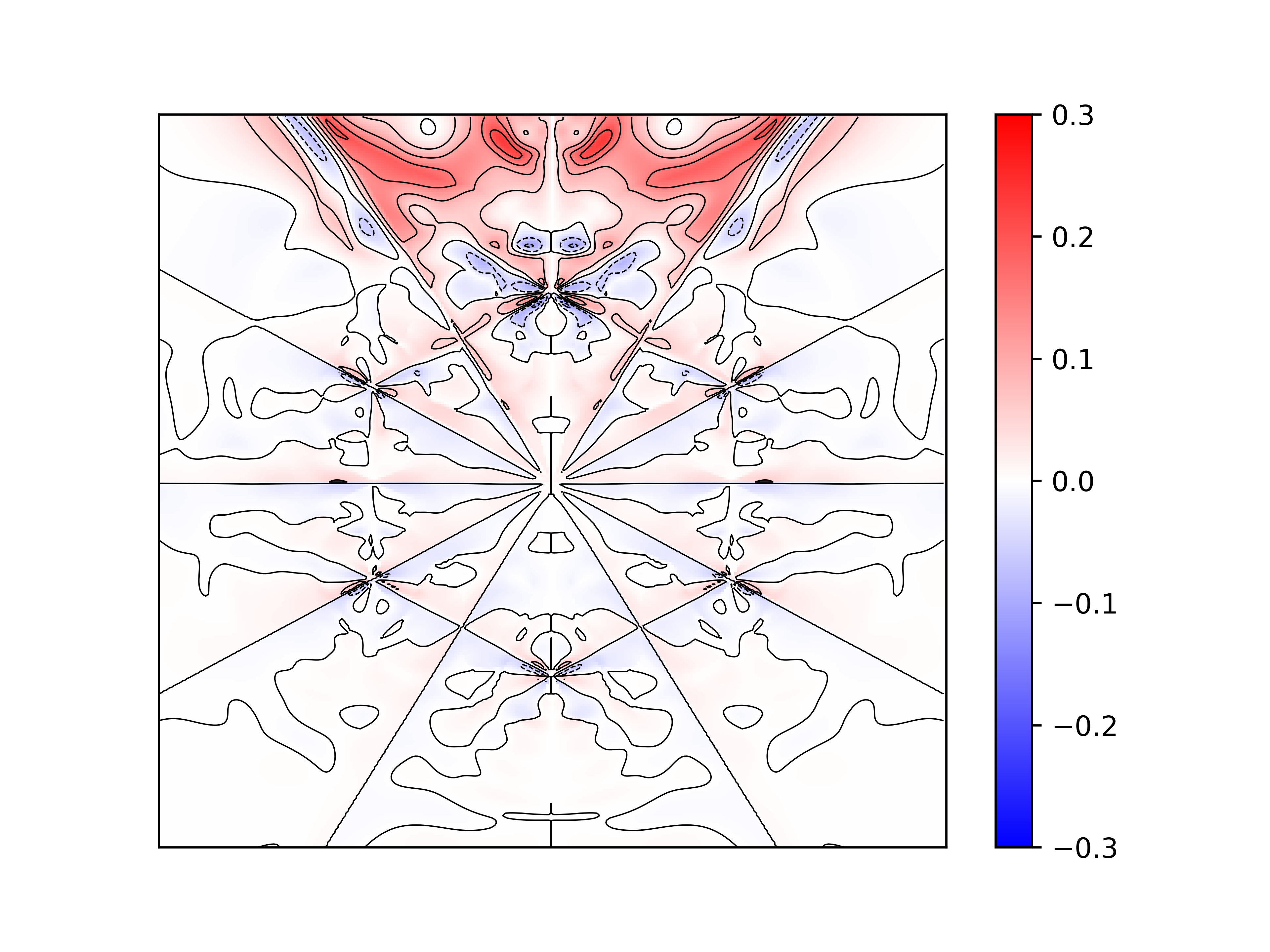}
    \caption{Difference in $C_2$ symmetry field upon mono-substitution with fluorine. Red regions (positive values) indicate a reduction in symmetry, while blue regions (negative values) indicate an increase in symmetry.}
    \label{fig:1f}
\end{figure}

\begin{figure}
    \centering
    \includegraphics[width=\linewidth]{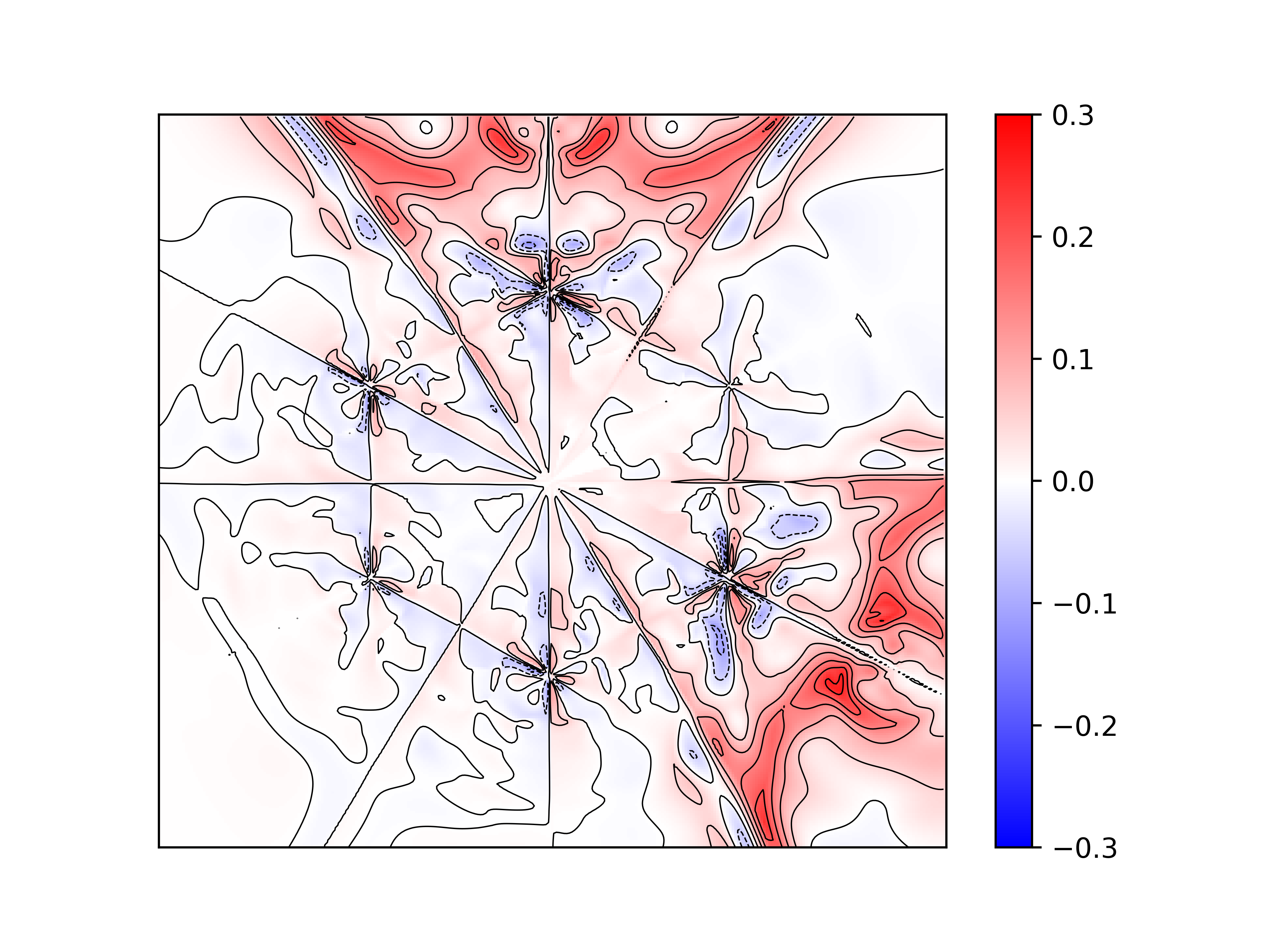}
    \caption{Difference in $C_2$ symmetry field upon 1,3-di-substitution with fluorine. Red regions (positive values) indicate a reduction in symmetry, while blue regions (negative values) indicate an increase in symmetry.}
    \label{fig:13f}
\end{figure}

Substituting a single fluorine onto the ring distorts the electron density around the benzene ring, and clearly destroys the local \ce{C-C} bond $C_2$ symmetry around the substituted carbon (\Cref{fig:1f}). There are, however, regions that exhibit an enhanced symmetry compared to benzene, indicated by the negative differences in blue color. This result suggests that the local density in fluorobenzene can establish distinctive quasi-$C_2$ symmetry which is not present in benzene. In addition, the two $C_2$ axes nearby the fluorine atom exhibit the largest distortion (a deflection of the quasi-symmetry axis), whereas other symmetry features in benzene are approximately preserved in fluorobenzene. Surprisingly, the largest changes in symmetry occur distant to the molecular frame, primarily constituting a (further) reduction in $C_2$ symmetry to either side of the conserved \ce{C-F} $C_2$ axis, but extending significantly perpendicular to the \ce{C-F} bond. These regions may indicate the imbalance in density between the \ce{C-F} bond and the axial fluorine lone pair. In 1,2-difluorobenzene, larger symmetry distortions are observed around all carbon atoms (Figure~S2). Nevertheless, the $C_2$ axis separating the two fluorine atoms remain unchanged, which is also observed in the model of 1,3-difluorobenzene  (\Cref{fig:13f}), and 1,2,3-difluorobenzene (Figure~S3). Interestingly, $C_2$ axes split the molecular environment into triangles that confine the symmetry malformation. The symmetry field of 1,3-difluorobenzene also shows that the \ce{C-H} bond between the two fluorine substitutions displays an absence of symmetry breaking, with symmetry distortion even less than the three other \ce{C-H} bonds, even at the C5 position which shares a conserved $C_2$ axis.

\subsection{Chirotopicity Field}

\begin{figure}
    \centering
    \includegraphics[width=\linewidth]{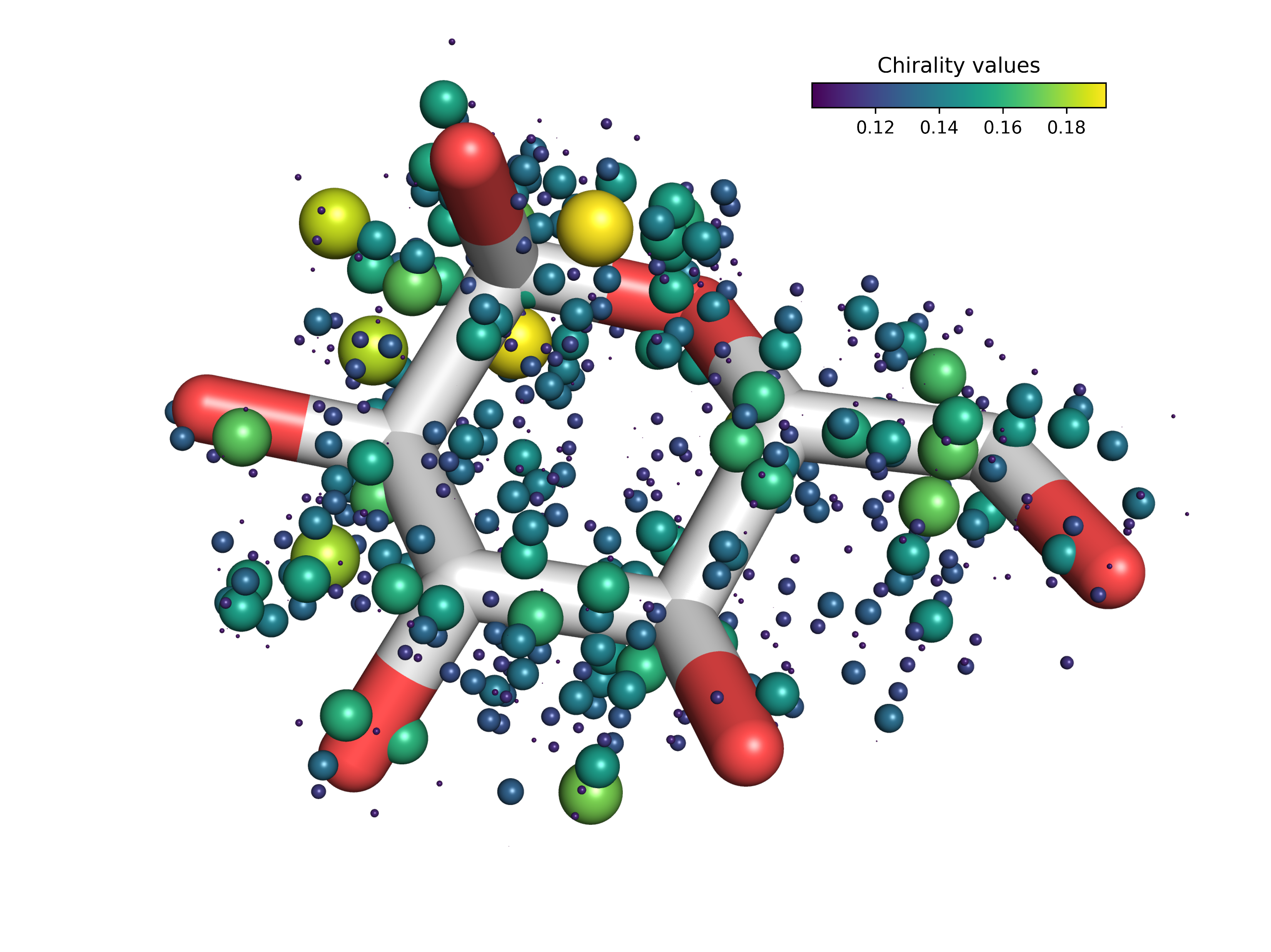}
    \caption{Chirotopicity field of D-glucose \textcolor{review}{($R=1$ a.u.)}. Larger, yellower spheres indicate larger values of $C_A$ and hence stronger local chiral environments.}
    \label{fig:glucose}
\end{figure}

We extend our previous symmetry analysis to a chiral representation of molecular systems. The traditional concept of an asymmetric (chiral) carbon is necessarily incomplete, as it neglects the influence of the surrounding environment. In fact, formally achiral atoms can contribute to the overall molecular chirality when they are embedded in a chiral environment, such as proximity to a chiral solvent, neighboring chiral atoms or functional groups, or an applied chiral external field. Consequently, the nature of the chiral environment surrounding a molecule is highly complex and remains underexplored. Within our framework, local chirality across the molecular region (i.e. the chirotopicity field) can be quantified at any point in space by evaluating the maximal deviation from any improper rotation symmetry based on local symmetry measures, \Cref{eq:chiral}.

\begin{figure}
    \centering
    \includegraphics[width=\linewidth]{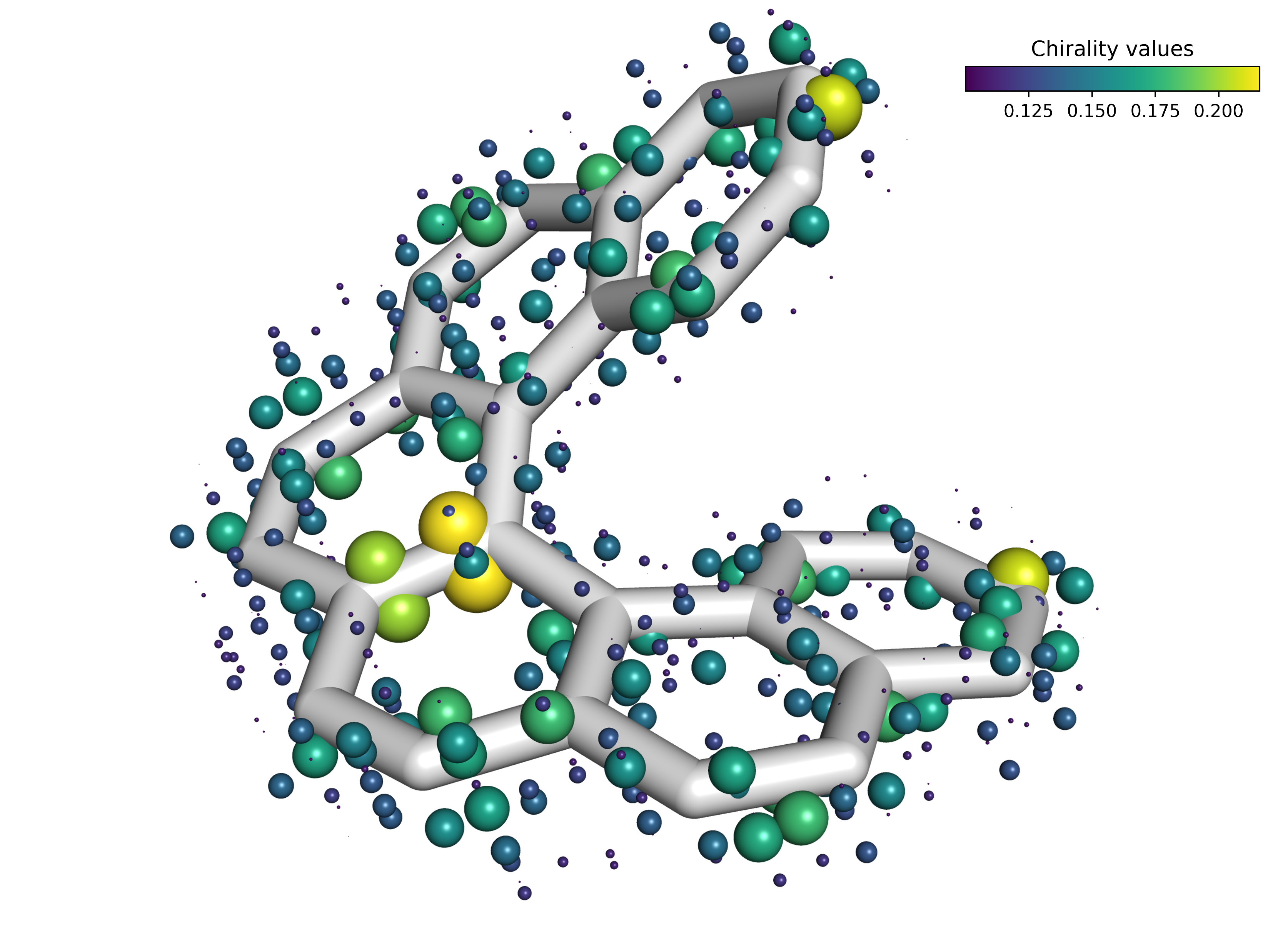}
    \caption{Chirotopicity field of hexahelicene \textcolor{review}{($R=1$ a.u.)}. Larger, yellower spheres indicate larger values of $C_A$ and hence stronger local chiral environments.}
    \label{fig:helicene}
\end{figure}

The chirotopicity fields of D-glucose and hexahelicene ([6]helicene) 
\textcolor{review}{computed at $R=1$ a.u.}
are shown in Figures~\ref{fig:glucose} and \ref{fig:helicene}, respectively. For visualization purposes, points with chirality values smaller than 0.1 are omitted. In this work, we consider the most stable chair conformer of glucose, in which all ring carbons (C1–C5) are stereogenic with equatorial hydroxyl groups, whereas the terminal carbon C6 is formally achiral. Owing to the presence of multiple stereogenic centers, the chiral environment of D-glucose is nonuniform and typically non-zero throughout the molecule. Regions of strong chirality are primarily distributed along the carbon skeleton, while the centroid of the pyranose ring exhibits negligible chirality. The most pronounced chiral region is localized around C1, which is strongly regulated by the bridging oxygen. Notably, despite being formally symmetric, carbon C6 displays substantial local chirality, arising from chirality transfer from the chiral ring framework.

In contrast to D-glucose, which contains multiple stereogenic centers, the chirality of hexahelicene originates intrinsically from its helical topology. As a result, its chiral environment differs significantly from that of D-glucose. As illustrated in \Cref{fig:helicene}, regions of significant chirality are distributed throughout the molecule, following the helical structure. However, the strongest chirality is observed in three distinct regions: the central twist and the tips of the two terminal rings. The enhanced chirality at the central rings indicates that these arenes experience the largest structural distortion. Meanwhile, the chiral enrichment at the terminal rings can be attributed to mutual inter-ring interactions. Because the dihedral angle between the terminal rings deviates significantly from planarity, the resulting torsional strain skews the local electron density, leading to enhanced chirality at the molecular termini.

\section{Conclusion}

In this work, we present a robust algorithm for quantifying local continuous symmetry in molecular systems. Our method has four-fold benefits:
\begin{enumerate}
\item The symmetry measure is formulated based on the electron density, allowing an accurate description of symmetry disorder arising from electronic perturbations induced by molecular interactions and external magnetic and electric fields. 
\item Local symmetry can be evaluated at any point in space within the molecular framework, without restriction to atomic positions. 
\textcolor{review}{\item The spatial extent of the analysis can be continuously tuned, allowing systematic probing of symmetry of atomic regions, chemical functional groups, molecular fragments, and subunits.}
\item The method has been implemented as a standalone Python package, which can be further integrated with existing quantum chemistry programs.
\end{enumerate}

However, the algorithm requires an extensive optimization procedure, which leads to a nontrivial computational cost. Moreover, because the analysis is based on the (relaxed) one-electron electron density obtained from self-consistent field (SCF) or correlated calculations, the results are inherently sensitive to the level of theory, basis set, convergence criteria, etc.

Overall, the proposed approach yields exact global symmetry measures while simultaneously revealing distinctive local symmetry and chirality features of molecules. 
\textcolor{review}{These local descriptors establish a connection between symmetry analysis and chemically observable phenomena by introducing a spatially resolved measure of symmetry in the electron density. By explicitly incorporating a controllable coverage scale, the approach reveals how symmetry is distributed across a molecule and how its local breaking correlates with functional regions such as reactive sites and interaction domains. This provides a foundation for linking symmetry-based descriptors with experimental observables, including stereoselectivity, intermolecular recognition, reactivity patterns, and spectroscopic response, as these depend on the anisotropy of the electronic environment.}
The framework is extendable to larger and more complex systems, such as coordination molecules, biomacromolecules, and condensed-phase environments, opening new avenues for exploring structure–property relationships through local symmetry and chirality.

\begin{acknowledgements}

This work was supported in part by the US National Science Foundation (grant CHE-2143725) and by the US Department of Energy (grant DE-SC0022893). Computational
resources for this research were provided by SMU’s O’Donnell Data Science and Research
Computing Institute.

\end{acknowledgements}

\section*{Conflict of Interest Statement}

The authors have no conflicts to disclose.

\section*{Data Availability Statement}

The data that support the findings of this study are available within the article and its supplementary material. Raw data is available from the
corresponding author upon reasonable request.

\bibliography{bib}

\end{document}